\documentclass[1p,12pt]{elsarticle}
\usepackage{amssymb,amsmath,color}
\usepackage{graphicx}
\usepackage{bm}
\usepackage{latexsym}
\usepackage{epsfig}
\usepackage[colorlinks]{hyperref} 
\journal{Astroparticle Physics}
\begin{document}
\begin{frontmatter}
\title{Gravitational Waves in Running Vacuum Cosmologies}

\author[dat]{D. A. Tamayo}
\ead{tamayo@if.usp.br}

\author[jasl]{J. A. S. Lima}
\ead{jas.lima@iag.usp.br}

\author[mesa]{M. E. S. Alves}
\ead{marcio.alves@ict.unesp.br}

\author[jcna]{J. C. N. de Araujo\corref{cor1}}
\ead{jcarlos.dearaujo@inpe.br}
\cortext[cor1]{Corresponding author}

\address[dat]{Instituto de F\'{i}sica, Universidade de S\~ao Paulo, Rua do Mat\~ao, 05508-090, S\~ao Paulo, SP, Brazil}
\address[jasl]{Departamento de Astronomia, Universidade de S\~ao Paulo, Rua do Mat\~ao 1226, 05508-900, S\~ao Paulo, Brazil}
\address[mesa]{Instituto de Ci\^encia e Tecnologia, Universidade Estadual Paulista,  S\~ao Jos\'e dos Campos, SP, 12247-016, Brazil}
\address[jcna]{Divis\~{a}o de Astrof\'{\i}sica, Instituto Nacional de Pesquisas Espaciais, \\ Avenida dos Astronautas 1758, S\~ao Jos\'e dos Campos, 12227-010 SP, Brazil}


\begin{abstract}
We investigate the cosmological production of gravitational waves  in a  nonsingular flat cosmology powered by a ``running vacuum" energy density { described by $\rho_{\Lambda}\equiv\rho_{\Lambda}(H)$, a phenomenological expression potentially linked with the renormalization group approach in quantum field theory in curved spacetimes.}   The model can be interpreted as a particular case of the class recently discussed by Perico et al. (Phys. Rev. D {\bf 88}, 063531, 2013) which is termed complete in the sense that the cosmic evolution  occurs between two extreme de Sitter stages (early and late time de Sitter phases). {The gravitational wave equation is derived and its time-dependent part numerically integrated since the primordial de Sitter stage. The generated spectrum of gravitons is also compared with the standard calculations where an abrupt transition, from the early de Sitter to the radiation phase, is usually assumed.} It is found  that the stochastic background of  gravitons is very similar to the one predicted by the cosmic concordance model plus inflation except at higher frequencies ($\nu \gtrsim 100$ kHz).  This remarkable signature of a ``running vacuum" cosmology  combined with the proposed  high frequency gravitational wave detectors {and measurements of the CMB polarization (B-modes) may provide a new window to confront more conventional models of inflation.} 
\end{abstract}


\end{frontmatter}
\section{Introduction}

In the last decade, many authors have proposed cosmological models driven by a { ``running vacuum'' energy density, $\rho_{\text{vac}}= \Lambda(H)/8\pi G$.} The leitmotiv of such an idea is related to two basic difficulties of the standard $\Lambda$CDM model (constant vacuum energy density). {Firstly,  a dynamical  $\Lambda(H)$-term} may solve the so-called cosmological constant problem because in this case the vacuum energy density relaxes to its present value and one may argue that $\Lambda$  is small today because the expanding Universe is too old. Secondly, a dynamical-$\Lambda(H)$ term may also solve the so-called cosmic coincidence problem, i.e. the fact that the time-varying matter energy density and the (constant) vacuum energy density have the same order of magnitude nowadays.

Since long ago, many different phenomenological decay laws for $\Lambda(H)$ were proposed in the literature (see \cite{Ozer1986, Carvalho1992, Lima1994, Lima1996, Overduin1998} for the oldest literature and \cite{N1,N1b,N2,N2b} for more recent articles). The predictions of the latest models have also been confronted with the available observational data and the results compared with the predictions of the standard $\Lambda$CDM cosmology \cite{Basilakos2009,Basilakos2012}. 

{Besides enlarging the standard view of a rigid $\Lambda$-term, there are also some attempts to justify theoretically a ``running vacuum" cosmology based on different methods, among them:  the thermal instability of a de Sitter spacetime \cite{Gasp} and the renormalization group (RG) approach in curved spacetimes \cite{Sola2011}.} In the latter case, for instance, the emerging dynamical $\Lambda(H)$-term  (beyond the true constant vacuum contribution) depends on an expansion power series on the Hubble parameter ($\sum H^{n}$) where only the even powers of H are involved in the RG realization, selected by the general covariance of the effective action appearing in the quantum field theoretical treatment  in curved spacetimes \cite{Sola2011,Shapiro2002,Shapiro2002b}.  { In the same vein, several alternative approaches} also try to represent the interacting $\Lambda(H)$ models through a Lagrangian description as in the $F(T)$ gravity approach \cite{Poplawski2006}, or in its generalized form $F(R,T)$, as discussed by Harko and coworkers \cite{Harko}. Other attempts involve  a mixture of a scalar field interacting with the radiation bath \cite{Maia2002,Bessada2013}, as in the so-called warm inflationary model \cite{Berera,Bererab,Bererac,Lima1999}, or still based on the quantum mechanics probability of unstable states \cite{Marek15}. Although interesting and highly promising to understand the decaying vacuum problem in the evolving Universe, none of them can at present be considered definitive and/or widely accepted by the community.  

In this connection, a large class of nonsingular models, where the vacuum energy density evolves phenomenologically as a truncated power-series in the Hubble parameter, has been proposed a couple of years ago \cite{Perico2013,Lima2013} (its dominant term $\rho_{\text{vac}}(H) \propto H^{n}, n > 2$). This class of models has some interesting features, among them: a nonsingular origin for the expanding universe (no horizon problem) with a deflationary process also without ``exit problem", that is, the model evolves smoothly from the primeval nonsingular de Sitter state to the radiation phase; its late-time cosmic expansion history is very close to the concordance model and it also furnishes a smooth link between the initial and final de Sitter stages through the radiation and matter dominated phases. The temperature behavior and the entropy generation during the continuous non-adiabatic  transition from de Sitter to the radiation phase has also been investigated \cite{Lima2015,Lima2015b} and a comparison with the late time observations has also been carried out in detail by Gomez-Valent and Sol\`a \cite{AS2015}.

{Furthermore, the recent LIGO detection of gravitational waves (GWs) opened a new window of observations for astronomy  and cosmological problems \cite{LIGO16}. Its importance goes much far beyond the search for signatures of  compact binary coalescence  (black holes, neutron stars, etc). For instance,  tensor perturbations describing GWS generated in the inflationary stage may affect the pattern of Cosmic Background Radiation (CMB) anisotropies through the B-modes polarization \cite{Planck2015}.
In addition, there are  also ongoing projects (BICEP3, Keck Array experiments, SPT) and future probes with high sensitivity  instruments like the QUBIC, especially designed  to measure the B-modes with high precision \cite{QUBIC, QUBICa}. Naturally,  as compared with more conventional inflationary models, the amplification of GWs  from a primordial de Sitter stage supported by a decaying-$\Lambda(H)$ model may affect CMB polarization in a different way, and, as such,  this line of inquire deserves a closer scrutiny.  

 In this context,  we analyze here the production of primordial GWs for a  class of nonsingular ``running vacuum cosmologies". The phenomenological $\Lambda(H)$-term  adopted here is defined by: $\Lambda(H) = \Lambda_b + \alpha H^{3}/H_I$}, where $\Lambda_{b}$ is the constant bare vacuum energy density, $H$ is the Hubble parameter, $H_I$ is the primordial inflationary scale, and $\alpha$ a dimensionless free parameter. Therefore, unlike the general class discussed in Refs. \cite{Perico2013,Lima2013}, the very late time behavior of the model analized here is exactly $\Lambda$CDM. The GW equation is derived and its time-dependent part numerically integrated since the primordial de Sitter stage. 
 
 As we shall see, for higher frequencies ($\nu \gtrsim 100$ kHz) the predicted spectrum departs from the standard inflationary prediction, and, as such, these models are distinguishable if high frequency GW detectors become operative and reach the expected sensitivity in the near future. In general grounds, this signal reinforces the possibility of a new observational approach to inflationary physics, and provides additional  motivation in the search for stochastic background of GWs at high frequencies \cite{Tashiro2004,Easther2006,Garcia2008}). It is also argued that similar results remain valid for generic decaying vacuum cosmologies with an initial de Sitter stage supported by a power-law $H^{n}$ with $n>2$. 

\section{The model: basic equations}\label{section cosmology}

\begin{figure*}[th]
\centerline{
\epsfig{figure=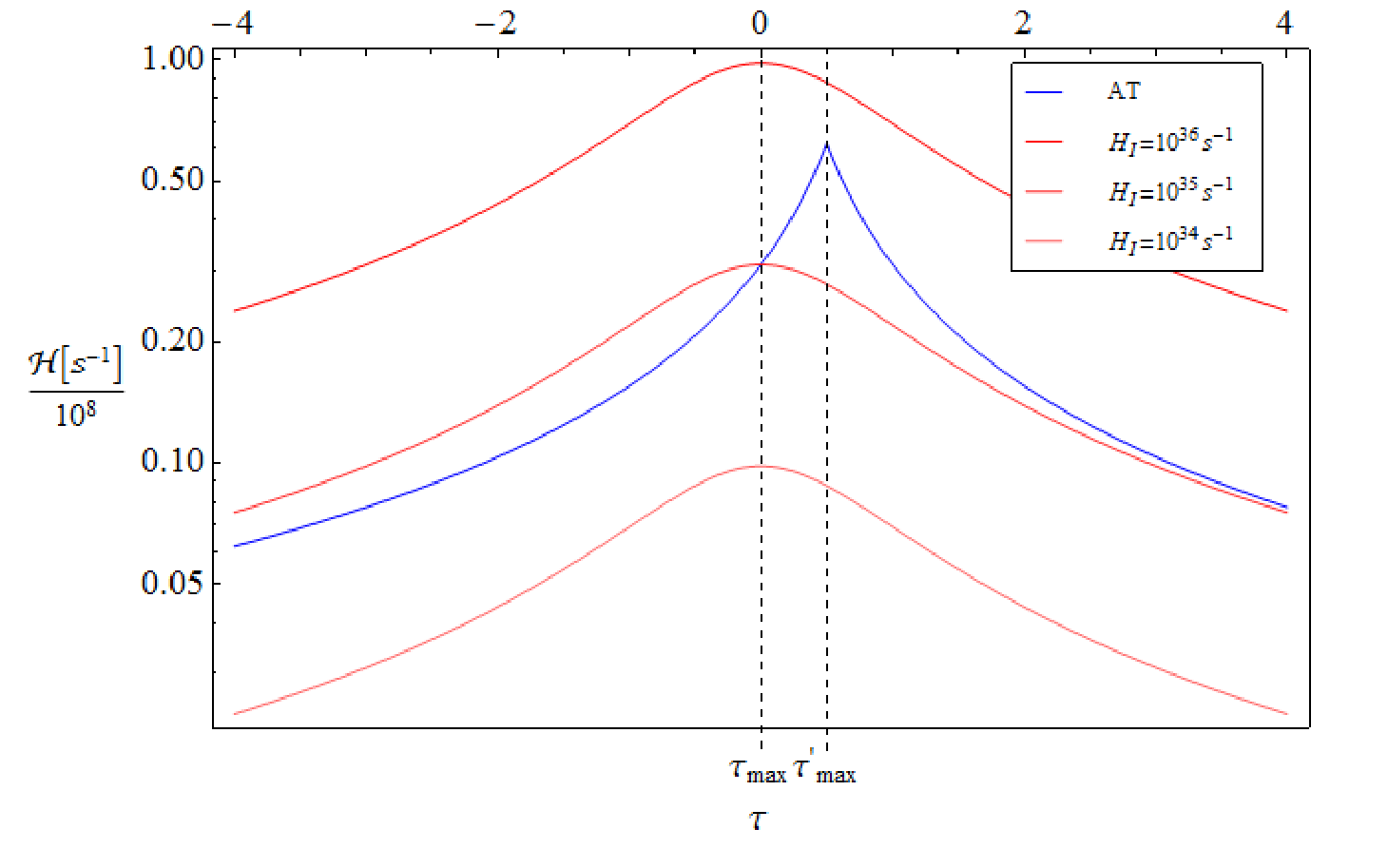,width=3.0truein}
\epsfig{figure=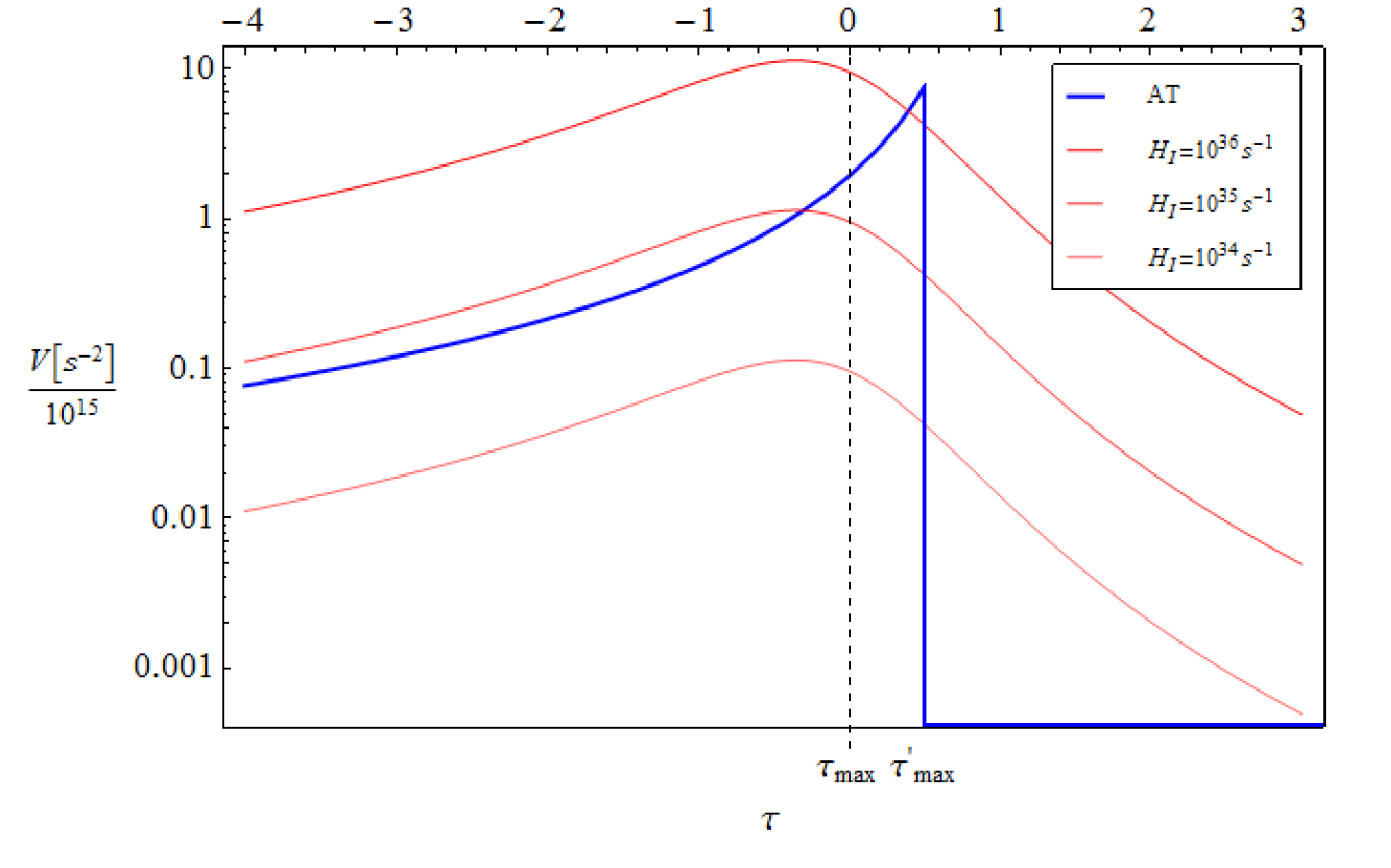,width=3.0truein}
\hskip 0.1in}
\caption{({\bf a}) Evolution of the quantity ${\cal H} = a'/a$ during the transition from primordial de-Sitter stage to the standard radiation phase ($\omega = 1/3$).  The blue line curve shows the evolution of ${\cal H}$ in the standard approach (inflation + radiation) assuming an abrupt transition (AT). The red lines display the evolution of  ${\cal H}$ in the smooth transition case (ST)  for different values of the initial Hubble parameter $H_I$  defining the early de Sitter phase. ({\bf b}) The same as Fig. 1{\bf a} but now for the potential, $V=a''/a$.   Again, the blue curve describes the ``Grishchuk potential'' usually  assumed in the AT treatment \cite{2Grishchuk1993,2Grishchuk1993b}. Note that the maximum height of the potential for the ST case is strongly dependent on the values of $H_I$. To better visualize the plots we have defined a suitable timescale $\tau(\eta)$ (see text). The time $\tau_{\text{max}'}$ for which the maximum value of  ${\cal H}$ and $V$ are attained assuming AT is always delayed in comparison to the ST case discussed here.}\label{fig1}
\end{figure*}
	
Let us now consider that the Universe is described by a flat Friedmann-Lem\^aitre-Robertson-Walker (FLRW) geometry. In the co-moving coordinate
system, the background line element reads ($c=1$):

\begin{equation} \label{eq1}
ds^2=dt^2-a^2(t)d\l^{2},
\end{equation}
where $a(t)$ is the scale factor.

{The Einstein equations in the above background read
	\begin{eqnarray}
	8\pi G\,\rho\,+\,\Lambda(t) &=& 3H^{2}\;,\label{eq5}\\
	8\pi\, G\, p - \Lambda(t) &=& -2{\dot H} - 3H^{2}\;,\label{eq6}
	\end{eqnarray} 
where a dot means time derivative and  $H = \dot a/a$ is the Hubble parameter}. To solve the above set of equations one needs the functional form of {$\Lambda(t)$, or equivalently, $\Lambda(H)$,}  as well as an equation of state (EoS).

As remarked before, it will be assumed here that the  $\Lambda(H)$-term  is given by a  particular case of the general class studied by Perico et al. \cite{Perico2013,Lima2013} (for closely related works see also \cite{Lima1994,Lima1996})

\begin{equation} \label{Eq4}
8 \pi G \rho_{\text{vac}}(H) =\Lambda(H) =  \Lambda_b + 3\alpha \frac {H^{3}}{H_I},
	\end{equation}
where $\Lambda_b$ is the bare cosmological constant.  By assuming that the material medium obeys the EoS, $p = \omega \rho$, where $\omega$ is a different constant for each era, one may show that the scale factor and the Hubble parameter obey the following equations (from now on, without loss of generality we consider $\alpha=1$):
\begin{eqnarray}\label{EM}
  2\dot{H} + 3(1 + \omega)H^2\left[1-\frac{H}{H_I}\right]- (1 + \omega)\Lambda_{b} = 0. \label{Eq5}
\end{eqnarray}
The standard cosmic concordance model equations are recovered by taking the limits  $H_I >> H$ and $\omega=0$ ($\Lambda$CDM). { At early times, the last term proportional to the bare $\Lambda_b$ can safely be neglected.} In this case,  for ({$\omega = 1/3$) the solution of the above equation takes the following form:

\begin{equation}
H = \frac{H_I}{1 + Ca^2}\,,
\end{equation}
where $C$ is an integration constant. We see that the transition from an early de Sitter ($Ca^{2} << 1,\, H\sim H_I$) to the radiation phase ($Ca^{2} >> 1, \,a \propto t^{1/2}$) is analytically described.

{ Now, in terms of the conformal time [$dt= a(\eta)d\eta$],  the line element (\ref{eq1})  becomes:}
\begin{equation}\label{C}
	ds^2=a^2(\eta)[d\eta^2 - \delta_{ij}dx^idx^j],
\end{equation}
while the equation of motion  (\ref{EM}) takes the form
\begin{eqnarray}\label{H_LH3_equation}
2\frac{H'}{a} + 3(1 + \omega)H^2\left[1-\frac{H}{H_I}\right]- (1 + \omega)\Lambda_{b} = 0,
\end{eqnarray}
where primes denote derivative with respect to $\eta$. By fixing the EoS parameters, the integration constants for each era are obtained by imposing the continuity conditions for $a(\eta)$ and $a^\prime(\eta)$ at the transition times between two subsequent eras.

By assuming that the vacuum decay mainly into ultra-relativistic particles ($\omega=1/3$) when the bare term of the decaying vacuum is negligible ($\Lambda_b << H^{3}/H_I$), it is easy to see that (\ref{H_LH3_equation}) boils down to $ a''- \frac{2}{H_I} \left(\frac{a'}{a}\right)^3 = 0$. A direct integration of this equation yields the following solution:

\begin{equation}\label{a_LH3_2}
   a(\eta) = \frac{1}{2C_1}\left[\eta+C_2 + \sqrt{(\eta+C_2)^2 + \frac{4 C_1}{H_I}}\right],
\end{equation}
which has two limiting cases: at very early times it is a de Sitter solution, $a\propto |\eta|^{-1}$, whereas at late time ($t >> H_I^{-1}$) , it enters in the radiation phase, $a\propto \eta$ \cite{Perico2013,Lima2015b}. The reduced Hubble parameter, $\mathcal{H}(\eta) = {a'}/{a}$, for this stage

\begin{equation}\label{reduc}
   \mathcal{H}_\text{Inf-R}(\eta) = \frac{a'}{a} = \left[(\eta + C_2)^2 + \frac{4 C_1}{H_I}\right]^{-1/2},
\end{equation}
assumes its maximum value for $\eta=-C_2$, so $
   \mathcal{H}(-C_2) = \mathcal{H}(\eta_{\text{max}}) \equiv \mathcal{H}_{\text{max}} = \sqrt{\frac{H_I}{4 C_1}}$. Therefore,  we can rewrite the integrating constants in the form $C_1 = 4 H_I^{-1} \mathcal{H}_{\text{max}}^2$ and $C_2 = -\eta_{\text{max}}$, and, naturally, (\ref{reduc}) can be  rewritten in terms of the new pair of constants ($\eta_{\text{max}},\mathcal{H}_{\text{max}}$), a form that will be useful in the next section.




\section{Cosmological tensor perturbations}

The classical tensor metric perturbation { for the conformal FLRW  flat metric (\ref{C})} can be written as \cite{Mukhanov}:

	\begin{equation}
	ds^2=a^2(\eta)[d\eta^2 - (\delta_{ij}+h_{ij})dx^idx^j],
	\end{equation}
where the perturbation $h_{ij}$ is small, $|h_{ij}|\ll 1$, is transverse-traceless and satisfy the gauge constraints: $h_{0\mu}=0,\, h^i_i=0, \,\nabla^jh_{ij}=0$.

 The first order evolution equation of $h_i^j$ is given by \cite{Mukhanov,Maia1996,TLB2015}
	\begin{equation}
	{h^{j}_i}''+2\frac{a'}{a}{h^{j}_i}'-\nabla^2 h^{j}_{i} = 16\pi Ga^{2}{\delta{\bar{T}}^{i}_{j}}_{(T)}=0 \label{heq},
	\end{equation}
where  in the last equality we used that ${\delta{\bar{T}}^{i}_{j}}_{(T)} \equiv 0$ because the total EMT (matter plus vacuum) has the perfect fluid isotropic form \cite{Mukhanov}. The general solution of the above equation can  be Fourier expanded in the standard way 
	\begin{eqnarray}\label{hij}
	h_{ij}(\eta,\textbf{x}) = \frac{\sqrt{16\pi G}}{(2\pi)^{3/2}}\int d^3\textbf{n}\displaystyle\sum_{r=+,\times} \overset{r}{ \epsilon}_{ij}	(\textbf{n})\nonumber\\
    \times\left[\overset{r}{h_n}(\eta)e^{i\textbf{n}\cdot\textbf{x}}\,\overset{r}{c}_{\textbf{n}} + \overset{r}{h_n^*}(\eta)e^{-i\textbf{n}	\cdot\textbf{x}}\,\overset{r}{c_{\textbf{n}}}^{\dag}\right],
 \end{eqnarray}
where $\overset{r}{h}_n(\eta)$ are the mode functions, $\textbf{n}$ is the comoving wave vector,  $\overset{r}{c_{\textbf{n}}}$ and $\overset{r}{c_{\textbf{n}}}^{\dag}$ are complex numbers, and $\overset{r}{\epsilon}_{ij}(\textbf{n})$  is the symmetric,  transverse-traceless polarization tensor.
Now, by recalling that the comoving wave number is given by

	\begin{equation}
	n=|\textbf{n}|=\frac{2 \pi a(\eta)}{\lambda}=k \,a(\eta),
	\end{equation}
and inserting the solution (\ref{hij}) into (\ref{heq}), it is readily seen that the temporal part yields a differential equation valid for both polarizations (henceforth we drop the index $r$)

	\begin{equation}
	h_n(\eta)''+2{\cal H(\eta)}h_n(\eta)'+n^2h_n(\eta)=0.
	\end{equation}
where we have conveniently used the reduced Hubble parameter, ${\cal H}$, whose continuity also guarantees the continuity of the first derivative of the scale factor. By  using the auxiliary function, $\mu = h_n(\eta)a(\eta)$, the above equation assumes the form first derived by Grishchuck \cite{2Grishchuk1993,2Grishchuk1993b}: 

	\begin{equation}\label{mu}
	\mu''+\left(n^2-\frac{a''}{a}\right)\mu=0,
	\end{equation}
This is the basic equation which allow us to obtain the associated  physical quantities like the wave amplitude, energy density and power spectrum.

The scale of the ``Grishchuck potential'', $V(\eta) = a''/a$, as compared with the wave-number determines the behavior of the limiting solutions for $\mu(\eta)$. The solutions are oscillatory, $\mu \propto e^{\pm in\eta}$, when $n^2 \gg |V|$ holds. The high-frequency modes in this case are diluted by the cosmic expansion since $h=e^{\pm in\eta}/a$. In the opposite regime one finds $\mu \propto a$ so that the amplitude remains constant. In this case, the damping of the waves due the universe expansion is avoided, a phenomenon usually  referred to as \textit{adiabatic amplification} \cite{2Grishchuk1993,2Grishchuk1993b}.

In Figures \ref{fig1}a and \ref{fig1}b we show the behavior of the quantities ${\cal H(\eta)}$  and  $V(\eta)$  during the transition from the early de Sitter stage to the radiation phase. For the sake of clarity, both quantities were expressed in terms of a convenient variable, $\tau={\cal H}_{\text{max}}(\eta_{\text{max}}-\eta)$,  where  $\eta_{\text{max}}$ is the time when ${\cal H}={\cal H}_{\rm max}$ is reached. As shown below Eq. (\ref{eqH}), the value of ${\cal H}_{\rm max}$ depends on the scale $H_I$. In the variable $\tau$, the maximum of ${\cal H}$ for the decaying vacuum model occurs  at $\tau_{\text{max}}=0$ while for the abrupt case  the maximum is delayed ($\tau'_{\text{max}}>\tau_{\text{max}})$. The interesting point here is that such a transition in our model is smooth and can analytically be followed in terms of the scale factor and the quantity ${\cal H}$.

\section{Gravitational wave solutions }
\begin{figure*}[th]
\centerline{
\epsfig{figure=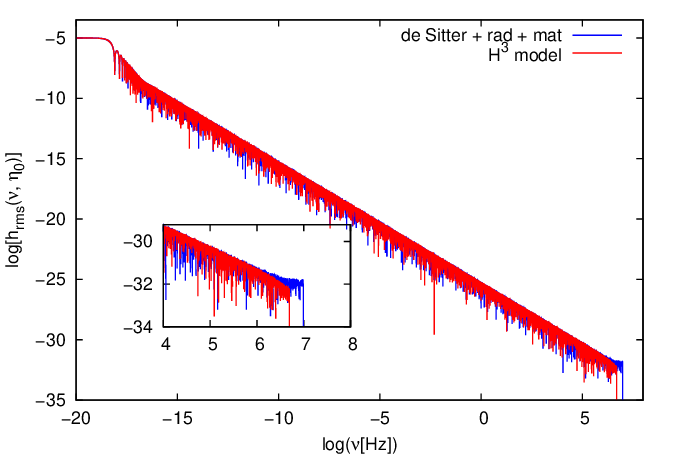,width=3.0truein}
\epsfig{figure=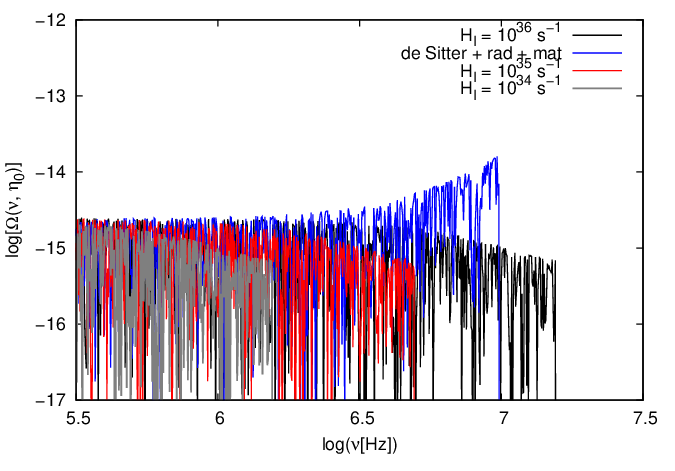,width=3.0truein}
\hskip 0.1in}
\caption{({\bf a}) Present day root-mean-square amplitude of the GWs as a function of the physical frequency for $H_I = 10^{35}~{\rm s}^{-1}$.  The amplitudes are displayed both for the decaying vacuum cosmology (red-curves) as well as for the $\Lambda$CDM model (blue-curves). In the former case the model evolves smoothly from inflation to radiation while in the latter an abrupt transition is assumed (see also Fig.1).  The superposition suggests that the predictions are quite similar in the domain of low frequencies. The infographic display the difference for high frequencies ($\nu \gtrsim 100$ kHz). ({\bf b}) Density parameter for GWs in the domain of high frequencies. For the $\Lambda$CDM model we have fixed the Hubble parameter to be $H_{inf}= 10^{35}~{\rm s}^{-1}$ while for the decaying vacuum cosmology the arbitrary scale $H_I$ assumed three possible values as indicated in the figure (see text).}\label{fig2}
\end{figure*}
Let us now determine the generated spectrum of the GWs. During the transition from inflation to radiation (Inf-R), the GW equation (\ref{mu}) assumes the form
\begin{equation}\label{eq_mu_inf}
\mu_{\text{Inf-R}}''+\left(n^2-V_\text{Inf-R}(\eta)\right) \mu_{\text{Inf-R}}=0,
\end{equation}
where $V_\text{Inf-R}(\eta)=a''/a \equiv {\cal H}_\text{Inf-R}^2 + {\cal H}^\prime_\text{Inf-R}$. From section \ref{section cosmology} one finds that the quantity ${\cal H}_\text{Inf-R}$ for $\omega = 1/3$ reads: 

\begin{equation}\label{eqH}
{\cal H}_\text{Inf-R}(\eta) = \frac{1}{\sqrt{(\eta - \eta_{\rm max})^2 + 1/{\cal H}_{\rm max}^2 }},
\end{equation}
where we have at $\eta_{\text{max}}$ the maximum value ${\cal H}_{\rm max} \simeq H_0 \sqrt{\eta_{\text{eq}}H_I/8}$ and $\eta_{\text{eq}}$ is the conformal time of equality of the energy densities of radiation and matter.  Since the scenario  starts as a de Sitter spacetime (see discussion below Eq. (6)),  it is possible to obtain an analytical solution for $\mu_{\text{Inf}}$ at that time. In addition, by imposing the adiabatic vacuum $\lim_{n \rightarrow \infty} \mu_{\text{Inf}} \propto e^{-in\eta}/ \sqrt{n}$ constraint [27], we find:

\begin{equation}
\mu_{\text{Inf}}(n,\eta)\simeq \frac{A_0}{\sqrt{2n}}\left(1-\frac{i}{n (\eta - \eta_{\text{max}})}\right) e^{-i n (\eta-\eta_{\text{max}})}.
\end{equation}
where $A_0$ is a real constant which specifies the initial amplitude of the GWs. The above expression for $\mu_{\text{Inf}}(n,\eta)$, as well as its derivative $\mu_{\text{Inf}}^\prime(n,\eta)$, provide the initial conditions (taken at the initial unstable de Sitter era) for the numerical integration of Eq. (\ref{eq_mu_inf}). { Consequently, the evolution of the GWs can be traced until a time $\eta_{\text{r}}$ from which the scale factor and waveforms behave as in the usual $\Lambda$CDM scenario. At the radiation era, it is possible to find an analytical solution for $\mu(\eta)$ whose initial conditions  (evaluated at $\eta_{\text{r}}$) are given by the previous numerical integration. This solution is valid until the conformal time $\eta_{\text{eq}}$, for which matter ($\omega = 0$) starts to dominate, and the scale factor evolves exactly as in the standard matter dominated cosmological scenario. At this stage, the analytical solution was obtained by considering the initial conditions (at $\eta_{\text{eq}}$) given at the end of the precedent radiation era. Finally, we have evaluated $\mu(\eta_0)$ at the present time $\eta_0$.} In addition, from the standard definition, we also find the resulting present power spectrum of the relic GWs in this model:
\begin{equation}
\mathcal{P}(n, \eta_0) = \frac{16\ell^2_{\rm Pl}}{\pi}n^3|h_n(\eta_0)|^2.
\end{equation}

In Figure \ref{fig2}a we show the present day root-mean-square (rms) amplitude of the GWs as a function of the physical frequency $\nu$ for the decaying vacuum model, which is related to the power spectrum via $h_{rms}(\nu,\eta_0) = \sqrt{\mathcal{P}(\nu, \eta_0)}$. For comparison, we also show the rms amplitude for the abrupt three phase transition model (de Sitter, radiation and matter) with $H_I = 10^{35}~{\rm s}^{-1}$. For simplicity, we have not included a late time $\Lambda_b$ dominated epoch  since the same effect for both models is obtained, namely, a little smaller value of $h_{rms}$ for the complete GW spectrum \cite{Miao2007}.

Notice the remarkable superposition of both spectra for almost the entire spectral range. The models are distinguishable only at very high frequencies, which are displayed in detail in Figure \ref{fig2}b, where we show the spectrum of the density parameter $\Omega_{\rm GW}(\nu, \eta_0)$. In this figure, we have fixed the value of $H_I$ for the abrupt transition model and considered some possible values of $H_I$ for the decaying vacuum cosmology.

At high frequencies the two models predict distinct spectra for a given value of $H_I$. As shown in Figure \ref{fig1}, such an effect can be understood in terms of the behavior of ${\cal H}(\eta)$ and $V(\eta)$. Since the  decaying vacuum model evolves smoothly from a de Sitter towards a radiation era, the shape of ${\cal H}(\eta)$ for the transition and the consequent lower value of ${\cal H}_{\text{max}}$ (for the same $H_I$) result in a lower high frequency GW production.

On the other hand, there is no adiabatic amplification for frequencies $\nu > {\cal H}_{\text{max}}$ so that we have introduced a cut-off at $\nu_{\text{max}} = {\cal H}_{\text{max}}$ in these graphs. For a given $H_I$ value, the cut-off frequency for the abrupt transition model is twice the cut-off frequency for the decaying vacuum model. However, since we have freedom to choose $H_I$ for the $\Lambda$(H)-model,  different cut-off frequencies are allowed (see Fig. \ref{fig2}b). In principle, even for different values of $H_I$, the shape of the curves predicted by the AT and ST models are quite different thereby suggesting a possible test for the underlying inflationary mechanisms.

\section{Final comments} 
We have investigated the production of GWs in the context of a flat nonsingular decaying vacuum cosmology. The dynamical $\Lambda$(H)-term was  phenomenologically modeled as $\Lambda (H) = \Lambda_{b} + H^3/H_I$.  This kind of model undergone a smooth transition from an early inflation to the standard radiation phase which can analytically  be described \cite{Lima1994,Lima1996,Perico2013,Lima2013}. Interestingly, the model explain the present day entropy content of the Universe stored in the relic radiation \cite{Lima2015,Lima2015b} without necessity of a highly non-adiabatic reheating stage. Such scenario is also free of horizon and ``graceful" exit problems, and is also in agreement with the observations at low and intermediate redshifts since it evolves to the standard $\Lambda_b$CDM cosmology. In comparison with the standard approach (big-bang, adiabatic inflation, reheating,  plus radiation and subsequente eras), the present study also provides a simple way to understand how a different scenario can modify the predictions concerning the generated spectrum of GWs.

Solutions for the GW equation were numerically obtained. The result furnishes a simple and definite example that details of the transition from an early de Sitter to the radiation phase plays an important role in the generation of the GW spectrum. The identified lower GW production for high frequencies ($\nu \gtrsim 100$ kHz) is a remarkable signature of the $\Lambda$ decay model (see Figs. \ref{fig1}b and \ref{fig2}b). Although far exceeding the frequencies (and sensitivity) of the existing GW detectors in ground-based  experiments and space-based successors like LISA, new kinds of interferometers and  detectors operating in a frequency range high enough to test primordial GW have been discussed in the recent literature \cite{Aktuso2008}(see also \cite{Arvanitaki2013,Tobar2014}). The high frequency behavior shows that the model is not only distinguishable from abrupt  inflationary scenario but can also provide a crucial test for the underlying mechanism as predicted in some multi-field inflationary models\cite{Easther2006}. Note also that current GW interferometers can not detect the putative GW background studied here, since the
amplitude of this background is orders of magnitude lower then the interferometer sensitivities. We refer the 
reader to the paper by Moore, Cole \& Berry \cite{Moore2015}, where the sensitivity curves of several detectors are displayed.
 Finally, it should be reinforced  that the model discussed here is a starting point for the investigation of more complex and rich decaying vacuum cosmologies, like the general class proposed in Ref. \cite{Perico2013,Lima2013}. Such models with $\Lambda(H) \propto H^{n}, n>2$ deserves a closer scrutiny since they furnish a complete cosmological history.  In principle, although the details might be dependent on the power $n$,  similar results to the ones derived here should  be expected because the entire class also evolves through a smooth transition from de Sitter to the radiation. { A detailed analysis involving the generation of GWs in this more general framework plus the implications in the observed pattern of CMB anisotropies through the B-modes polarization will be discussed in a forthcoming communication.}

\section*{Acknowledgements}
DT is supported by a fellowship from CAPES. JASL and JCNA would like to thank CNPq, FAPESP and CAPES (PROCAD2013) for partial financial support. We would also like to thank the referees for their helpful comments and suggestions.


\begin{thebibliography}{00}
\bibitem{Ozer1986} M. Ozer and M. O. Taha, Phys. Lett.  B 171 (1986) 363.
\bibitem{Carvalho1992} J. C. Carvalho, J. A. S. Lima and I. Waga, Phys. Rev. D 46 (1992) 2404.
\bibitem{Lima1994} J. A. S. Lima and J. M. F. Maia, Phys. Rev. D 49 (1994) 5597.
\bibitem{Lima1996} J. A. S. Lima and M. Trodden, Phys. Rev.  D 53 (1996) 4280, arXiv: astro-ph/9508049.      
\bibitem{Overduin1998} J. M. Overduin and F. I. Cooperstock, Phys. Rev. D 58 (1998) 043506.
\bibitem{N1} S. Carneiro and J. A. S. Lima, Int. J. Mod. Phys. A 20 (2005) 2465, arXiv: gr-qc/0405141. 
\bibitem{N1b} J. S. Alcaniz and J. A. S. Lima, Phys. Rev. D 72 (2005) 063516, arXiv: astro-ph/0507372.  
\bibitem{N2} H. Borges, S. Carneiro, J. Fabris, C. Pigozzo, Phys. Rev. D 77 (2008) 043513.
\bibitem{N2b} F. E. M. Costa, J. A. S. Lima and F. A. Oliveira, Class. Quantum Gravity  31 (2014) 045004. 
\bibitem{Basilakos2009} S. Basilakos, Mon. Not. R. Astron. Soc. 395 (2009) 234.
\bibitem{Basilakos2012} S. Basilakos, D. Polarski and J. Sol\`a, Phys. Rev. D 86 (2012) 043010.
\bibitem{Gasp} M. Gasperini, Phys. Lett. B 194 (1987) 347.
\bibitem{Sola2011} J. Sol\`a, J. Phys. Conf. Ser. 283 (2011) 012033.
\bibitem{Shapiro2002} I. L. Shapiro and  J. Sol\`a,  Phys. Lett. B 475 (2000) 236.
\bibitem{Shapiro2002b} I. L. Shapiro and  J. Sol\`a, J. High Energy Phys.  02 (2002) 006, arXiv: hep-th/0012227.  

\bibitem{Poplawski2006} N. J. Poplawski, arXiv: gr-qc/0608031v2.
\bibitem{Harko}  T. Harko, F. S. N. Lobo, S. Nojiri, S. D. Odintsov, Phys.Rev. D 84, 024020 (2011), arXiv:1104.2669 
\bibitem{Maia2002} J. M. F. Maia and J. A. S. Lima, Phys. Rev.  D 65 (2002) 083513, arXiv: astro-ph/0112091.
\bibitem{Bessada2013} D. F. A. Bessada, Phys. Rev. D 88, (2013) 023005, arXiv: 1307.1099 [gr-qc]
\bibitem{Berera} A. Berera and L. Z. Fang, Phys. Rev. Lett. {74} (1995) 1912 .        
\bibitem{Bererab} A. Berera, Phys. Rev. Lett. 75 (1995) 3218. 
\bibitem{Bererac} A. Berera, I. Moss and R. Ramos,  Rept. Prog. Phys. 72 (2009) 026901. 
\bibitem{Lima1999} J. M. F. Maia and J. A. S. Lima, Phys. Rev.  D 60 (1999) 101301, arXiv: astro-ph/9910568. 
\bibitem{Marek15} M. Szydlowski, Phys.Rev. D 91, 123538 (2015), arXiv:1507.02114 
\bibitem{Perico2013} E. L. D. Perico, J. A. S. Lima, S. Basilakos and J. Sol\`a, Phys. Rev. {D88} (2013) 063531, arXiv:1306.0591 [astro-ph.CO]. 
\bibitem{Lima2013} J. A. S. Lima, S. Basilakos  and J.  Sol\'a, Mon. Not. R. Astron. Soc. 431 (2013) 923, arXiv:1209.2802 [gr-qc].
\bibitem{Lima2015} J. A. S. Lima, S. Basilakos  and J.  Sol\'a,  Gen. Rel. Grav. 47 (2015) 40 arXiv:1412.5196 [gr-qc]. 
\bibitem{Lima2015b}  J. A. S. Lima, S. Basilakos and J. Sol\`a, Eur. Phys. J. { C76} (2016) 228, 1509.00163v2. 
\bibitem{AS2015} A. Gomez-Valent and  J. Sol\`a, Mon. Not. R. Astron. Soc. 448  (2015) 2810.
\bibitem{LIGO16} B. Abbott et al., Phys. Rev. Lett. { 116} (2016) 061102 
\bibitem{Planck2015} P. A. R. Ade et al. (Planck Collaboration) XIII. Cosmological Parameters, arXiv:1502.01589 [astro-ph.CO] (2015).
\bibitem{QUBIC} E. Battisteli et al., The QUBIC collaboration (2010), arXiv:10100645 (2010).
\bibitem{QUBICa} A. Ghrib et al., ``{\it Latest Progress on the QUBIC Instrument}", J. Low Temp. Phys. 176 (2014) 698 
\bibitem{Tashiro2004} H. Tashiro, T. Chiba and M. Sasaki, Class. Quant. Grav. 21 (2004) 1761.
\bibitem{Easther2006} R. Easther and E. A. Lim, J. Cosmol. Astropart. Phys.  04 (2006) 010.
\bibitem{Garcia2008} J. Garcia-Bellido, D. G. Figueroa and A. Sastre, Phys. Rev.  D 77 (2008) 043517.
\bibitem{Maia1996} M. R. de Garcia Maia and J. A. S. Lima, Phys. Rev. D 54 (1996) 6111.
\bibitem{TLB2015} D. A. Tamayo, J. A. S. Lima and D. Bessada, arXiv:1503.06110 [astro-ph.CO].
\bibitem{Mukhanov} V. Mukhanov, Physical Foundations of Cosmology,  Cambridge University Press, 2010.
\bibitem{2Grishchuk1993}   L. P. Grishchuk,  J. Experimental Theoretical Phys. 40, 409 (1975).
\bibitem{2Grishchuk1993b} L. P. Grishchuk, Class. Quant. Grav. { 10} (1993) 2449 .  
\bibitem{Miao2007} H.X. Miao and Y. Zhang, Phys. Rev. { D75}, (2007) 104009.
\bibitem{Aktuso2008} T.~Akutsu {\it et al.},  Phys. Rev. Lett. 101 (2008) 101101, arXiv:0803.4094 [gr-qc]
\bibitem{Arvanitaki2013} A. Arvanitaki and A. A. Geraci, Phys. Rev. Lett. 110 (2013) 071105. 
\bibitem{Tobar2014} M. Goryachev and M. E. Tobar, Phys. Rev. D 90 (2014) 102005. 
\bibitem{Moore2015} C. J. Moore, R. H. Cole and  C. P. L. Berry, Class. Quant. Grav. {32} (2015) 015014 .
\end{thebibliography}
\end{document}